\documentclass[12pt, preprint]{aastex}

\usepackage{natbib}

\newcommand{\rhos}{\rho_{\star}}
\newcommand{\Os}{\Omega_{\star}}
\newcommand{\Om}{\Omega_{\rm M}}
\newcommand{\Ol}{\Omega_{\Lambda}}
\newcommand{\Ob}{\Omega_b}

\newcommand{\Lbox}{L_{\rm box}}

\newcommand{\Lam}{\Lambda}

\newcommand{\Del}{\Delta}
\newcommand{\hinv}{{h^{-1}}}

\newcommand{\himpc}{\hinv{\rm\,Mpc}}
\newcommand{\hikpc}{\hinv{\rm\,kpc}}
\newcommand{\kms}{{\rm\,km\ s^{-1}}}

\newcommand{\yr}{{\rm yr}}
\newcommand{\Msun}{M_{\odot}}
\newcommand{\himsun}{\hinv{\Msun}}
\newcommand{\ltsim}{\lesssim}
\newcommand{\gtsim}{\gtrsim}
\slugcomment{DRAFT VERSION}

\shorttitle{Is There a Missing Galaxy Problem at High Redshift?}
\shortauthors{Nagamine et al.}

\begin{document}

\title{Is There a Missing Galaxy Problem at High Redshift?}

\author{Kentaro Nagamine\altaffilmark{1}, Renyue Cen\altaffilmark{2}, Lars Hernquist\altaffilmark{1}, Jeremiah P. Ostriker\altaffilmark{2,3},\\ \& Volker Springel\altaffilmark{4}}

%\affil{Harvard-Smithsonian Center for Astrophysics, 60 Garden Street, Cambridge, MA 02138}
%\email{knagamine@cfa.harvard.edu}

\altaffiltext{1}{Harvard-Smithsonian Center for Astrophysics, 
60 Garden Street, Cambridge, MA 02138, U.S.A. \\
Email: knagamin@cfa.harvard.edu}

\altaffiltext{2}{Princeton University Observatory, Princeton, NJ 08544, U.S.A.}

\altaffiltext{3}{Institute of Astronomy, University of Cambridge, Madingley Road, 
Cambridge, CB3, OHA, UK}

\altaffiltext{4}{Max-Planck-Institut f\"{u}r Astrophysik, 
        Karl-Schwarzschild-Stra\ss{}e 1, 85740 Garching bei 
        M\"{u}nchen, Germany}

%%%%%%%%%%%%%%%%%%%%%%%%%%%%%%%%%%%%%%%%%%%%%%%%%%%%%%%%%%%%%%%%%%%%%%

\begin{abstract}
We study the evolution of the global stellar mass density in a $\Lam$
cold dark matter ($\Lam$CDM) universe using two different types of
hydrodynamical simulations (Eulerian TVD and SPH) and the analytical
model of \citet{Her03}.  We find that the theoretical calculations all
predict both a higher stellar mass density at $z\sim3$ than indicated
by current observations, and that the peak of the cosmic star
formation rate history should lie at $z\gtsim 5$.  Such a star
formation history implies that as much as (70\%, 30\%) of the total
stellar mass density today must already have formed by $z=(1, 3)$.  Our
results suggest that current observations at $z\sim 3$ are missing as
much as 50\% of the total stellar mass density in the Universe,
perhaps owing to an inadequate allowance for dust obscuration in
star-forming galaxies, limited sample sizes, or cosmic variance.
We also compare our results with some of the updated semi-analytic 
models of galaxy formation.
\end{abstract}

\keywords{cosmology: theory --- stars: formation --- 
galaxies: formation --- galaxies: evolution --- methods: numerical}

%%%%%%%%%%%%%%%%%%%%%%%%%%%%%%%%%%%%%%%%%%%%%%%%%%%%%%%%%%%%%%%%%%%%%%

\section{Introduction}
\label{section:intro}

Is the evidence for high redshift galaxy formation consistent with the
concordance $\Lam$CDM model?  Recent observational results include the
discovery of Extremely Red Objects at $z\ge 1$ \citep[e.g.][]{Elston88, 
McCarthy92, Hu94, Cimatti03, Smail02}, Sub-millimeter galaxies at 
$z\ge 2$ \citep[e.g.][]{Smail97, Chapman}, Lyman-break galaxies (LBGs) 
at $z\sim 3$ \citep[e.g.][]{Steidel99}, and galaxies at $z\gtsim 4$ either 
by their Lyman-$\alpha$ emission \citep[e.g.][]{Hu99, Rhoads, 
Taniguchi, Kodaira, Ouchi03a} or by their optical to near infrared (IR) 
colors \citep[e.g.][]{Iwata, Ouchi03b, Dick03b}.

Multiband photometry including the near-IR band makes it possible to
estimate the stellar mass of these high redshift galaxies by fitting
the observed photometric results with artificial galaxy spectra 
generated by a population synthesis model.  Using this technique,
several groups have now estimated the stellar mass density in the
Universe in the redshift range of $0\le z \le 3$ \citep[e.g.][]{Brinch,
Cole, Cohen, Dick03a, Fontana03, Glazebrook04}. \citet{Rudnick03} also 
estimated the 
stellar mass density at $z=0-3$ by combining the estimates of the 
rest-frame optical luminosity density and the mean cosmic mass-to-light 
ratio. These observational estimates constrain the evolution of the 
stellar mass density $\Os$ as a function of redshift or cosmic time.

The observations indicate the presence of a significant stellar
population at high redshift, and, by comparing with semi-analytic
models of galaxy formation, some authors claim that $\Lam$CDM models 
seriously {\it underpredict} galaxy formation at $z\sim 3$.
For example, \citet{Fontana03} compare their estimate of $\Os$ from the
Hubble Deep Field (HDF) South with the semi-analytic model of
\citet{Menci02}, and argue that the high-mass tail of the galaxy stellar
mass function is not adequately described by CDM models.  Comparing to
the same semi-analytic model, \citet{Poli} argue that hierarchical
models lack sufficient star formation at $z=2-4$, resulting in a
failure to reproduce the pronounced brightening of the luminosity
function at these redshifts.  Also, \citet{Dick03a} find 
that their data from the HDF North suggest a steeper increase in $\Os$ 
than some semi-analytic models \citep[]{Kau99, Som, Cole00}, and 
some of the semi-analytic models predict {\it higher} stellar mass 
density compared to their data points. 
Given these contradicting claims and the large parameter space 
available to the semi-analytic models as well as the limitations
of the current observational samples, it is not clear at present 
if any of these discrepancies actually pose a serious problem to 
hierarchical evolution.

In this paper, we compare the observational data with the results from 
state-of-the-art cosmological simulations of the standard concordance 
$\Lambda$CDM model and the theoretical model of \citet{Her03} 
(hereafter H\&S model), to show that, contrary to some of the claims, 
theory predicts a {\it higher} $\Os$ at $z=3$ than indicated by current 
observations and 
that the cosmic star formation rate (SFR) density peaks at $z\geq 5$, 
earlier than suggested by most semi-analytic models. Also, we will 
explicitly compare our results with those from the updated 
semi-analytic models by \citet{Som}, \citet{Granato00} (GALFORM), and 
\citet{Menci02}.

%%%%%%%%%%%%%%%%%%%%%%%%%%%%%%%%%%%%%%%%%%%%%%%%%%%%%%%%%%%%%%%%%%%%%%

\section{Simulations}
\label{sec:simulation}

We will show results from two different types of cosmological 
hydrodynamic simulations.  Both approaches include ``standard'' 
physics such as radiative cooling/heating, star formation,
and supernova (SN) feedback, although the details of the models and
the parameter choices differ somewhat.

One set of simulations was performed using an Eulerian approach, which
relies on a particle-mesh method for the gravity and the Total
Variation Diminishing (TVD) method \citep{Ryu93} with a fixed mesh for
the hydrodynamics.  The treatment of the radiative cooling and heating
is described in \citet{Cen92} in detail. The structure of the code is
similar to that of \citet{CO92, CO93}, but the code has significantly 
improved over the years with additional input physics. It has been 
used for a variety of studies, including the evolution of the 
intergalactic medium \citep{CO94, CO99a, CO99b}, damped 
Lyman-$\alpha$ absorbers \citep{Cen03}, and galaxy formation 
\citep*[e.g.][]{CO00, Nag01a, Nag01b, Nag02}.

Our other simulations were done using the Lagrangian Smoothed Particle 
Hydrodynamics (SPH) technique.  We use an updated version of 
{\small{GADGET}} \citep{Gadget}, which uses an `entropy conserving' 
formulation \citep{SH02} to mitigate problems with energy/entropy 
conservation \citep[e.g.][]{Her93} and overcooling.  This code also 
uses a multiphase model  of the interstellar medium to describe
self-regulated star formation and a phenomenological model for galactic 
winds \citep{SH03a}.  This approach has been used to study 
the evolution of the cosmic SFR \citep{SH03b}, damped 
Lyman-$\alpha$ absorbers \citep*{NSH04a, NSH04b}, and 
galaxies at high redshifts \citep*{NSHM}.

The cosmological parameters adopted in the simulations are intended to
be consistent with recent observational determinations 
\citep[e.g.][]{Spergel03}, as summarized in 
Table~\ref{table:simulation}.

%%%%%%%%%%%%%%%%%%%%%%%%%%%%%%%%%%%%%%%%%%%%%%%%%%%%%%%%%%%%%%%%%%%%%%

\section{Stellar Mass Density}
\label{sec:omegastar}

Figure 1 shows the evolution of the global stellar mass 
density $\Os$ as a function of redshift.
Figure~1a compares the simulation results with the observations, 
and Figure~1b compares the semi-analytic models with the observations.
Our results are scaled to unity at $z=0$ (see discussion below).
Observational data points are normalized to the local estimate by 
\citet[][filled circle at z=0]{Cole} following \citet{Rudnick03}.
The result of the H\&S model shown in both panels is simply the 
integral of their approximate formula for the SFR which will be 
given in Equation~(\ref{eq:sfr}).

For recent epochs ($z\lesssim 1$), the computed results are within the
range of the observational data. However, the important result here is
that at $1\ltsim z\ltsim 3$, all the observational estimates 
(both corrected and uncorrected estimates for the incompleteness of 
the survey) are smaller than the simulation results by more than a 
factor of two. The early development of the stellar mass density is much 
faster in the simulations than is suggested by current observations.  
As we discuss in Section~\ref{sec:sfr}, the larger $\Os$ at $z\sim 3$ 
originates from a higher SFR at $z\gtsim 3$ which peaks at 
$z\gtsim 5$ in both the TVD and SPH simulations.  Such a star 
formation history implies that about 70\% (50, 30, 15\%) of the 
total stellar mass density today must have been in place by 
$z=1$ ($z=2, 3, 5$). The contribution to $\Os(z=0)$ from the star
formation at $z>6$ is about 10\%.

We note that the data points of 
\citet[][magenta filled squares]{Rudnick03} should be considered
as a lower limit, as they did not attempt to make any extrapolations 
to correct for the incompleteness of their data. On the other hand, 
\citet[][green square boxes]{Dick03a} correct for the incompleteness 
of their data by integrating the Schechter fit with a faint-end slope
of $\alpha=-1.4$ down to a fainter magnitude.
For the \citet[][blue crosses]{Fontana03} data points, we have 
plotted the values for the SMC extinction case (which yields slightly 
larger $\Os$ values compared to the \citet{Calzetti00} extinction), 
and the upper error bars are extended up to their `maximum mass' 
estimates. 
For the \citet[][black filled triangles]{Glazebrook04} data points, 
we plotted the values for the mass limit of $\log(M/\Msun)=10.2$ which 
yields largest $\Os$ estimates from their data. We also note that 
dust extinction is taken into account in the analyses by 
\citet{Brinch, Dick03a, Fontana03, Glazebrook04} by allowing the 
extinction parameter to vary when fitting the broadband photometric 
measurements of each galaxy by the spectral energy distribution 
generated by a population synthesis model. \citet{Rudnick03} adopt 
the Calzetti extinction with $E(B-V)=0.30$ when deriving the mean 
mass-to-light ratio from the rest-frame $U-V$ color.

\begin{figure}
\epsscale{1.1}
\plottwo{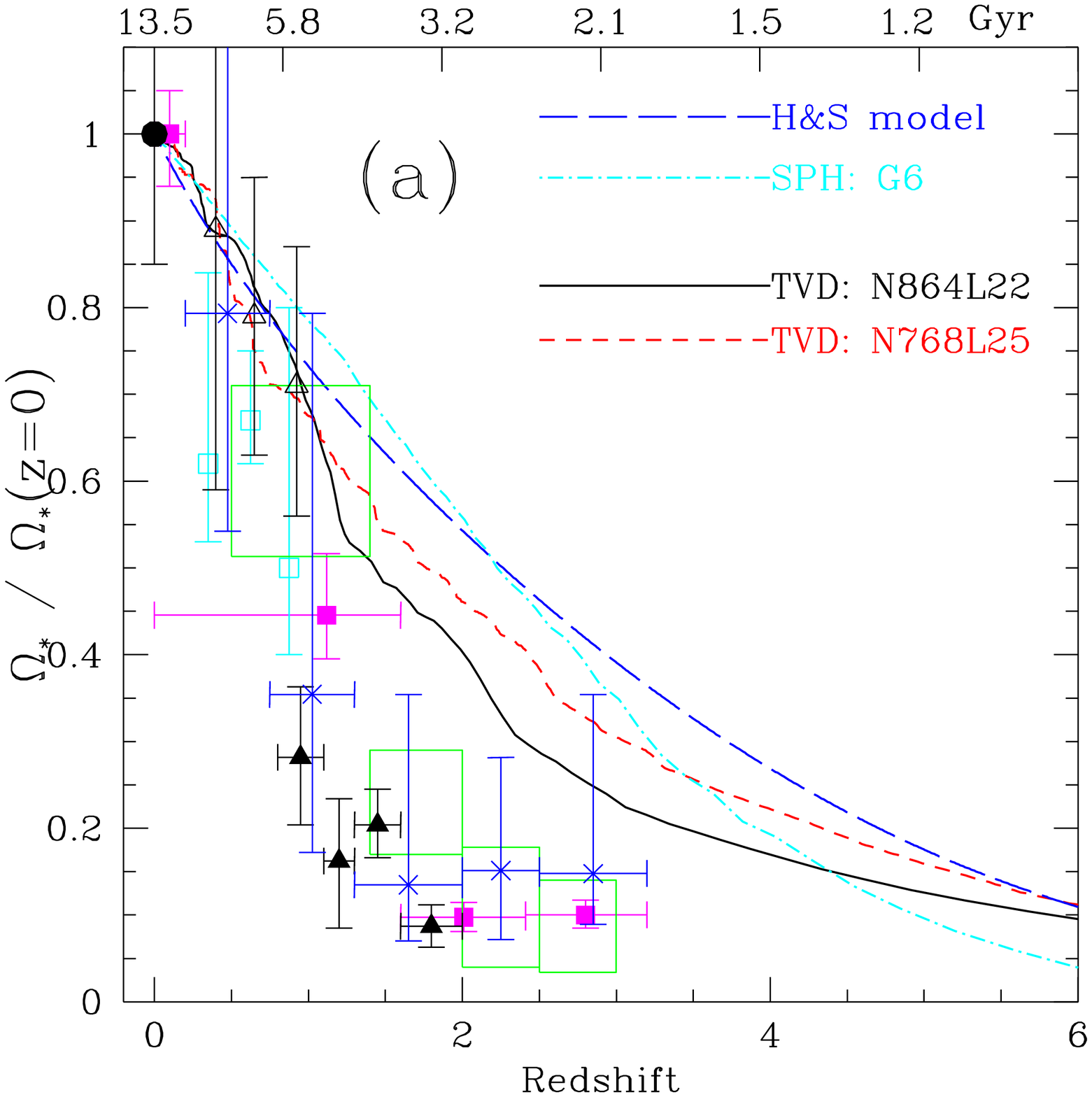}{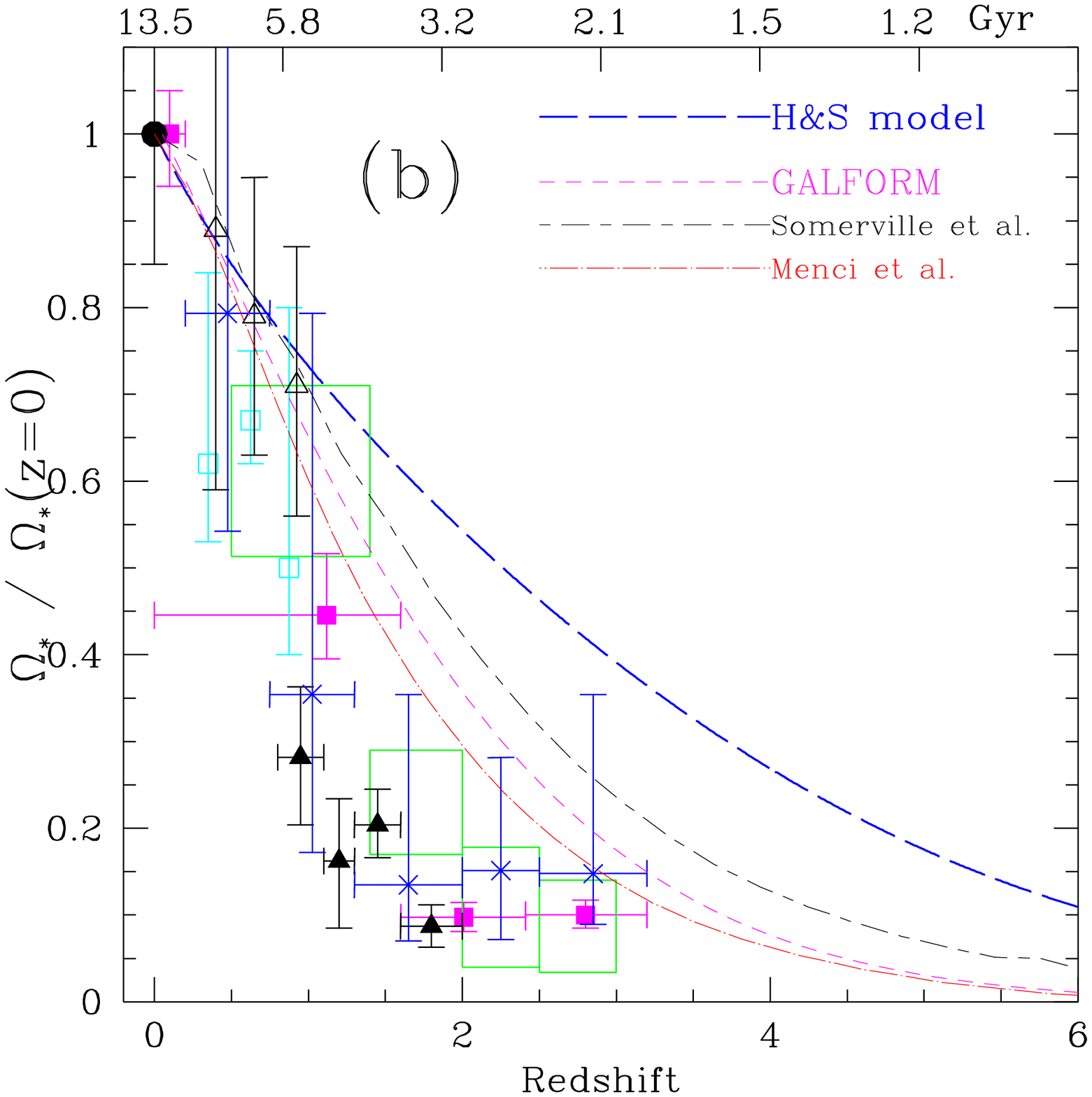}
\caption{Development of stellar mass density as a function of redshift.
{\it Panel (a)}: The black solid line is from the most recent TVD 
N864L22 run (consistent with the not shown N864L11 run), and the 
red short-dashed line is from the TVD N768L25 run which was used 
in \citet{Nag01a}. 
The blue long-dashed line is the analytical model of \citet{Her03},
and the cyan dot-long-dashed line is the result from the SPH `G6'-run.
The observational data points are, from low to high redshift,  
\citet[][black filled circle at $z=0$]{Cole}, 
\citet[][cyan open squares at $z=0.5 - 1$]{Brinch}, 
\citet[][black open triangles at $z=0.5 - 1$]{Cohen}, 
\citet[][black filled triangles at $z=1-2$]{Glazebrook04}, 
\citet[][magenta filled squares at $z=0-3$]{Rudnick03}, and 
\citet[][blue open crosses at $z=0.5-3$]{Fontana03}. 
The four green boxes are from \citet{Dick03a} which show the range 
of systematic uncertainty introduced by varying the metallicity and 
star formation histories of the mass-fitting model they used to 
derive the stellar mass density. All the data points are normalized 
to the local estimate by \citet{Cole} following \citet{Rudnick03}.
{\it Panel (b)}: Observational data points and the H\&S model are 
the same as panel (a). The results of two semi-analytic models by 
\citet[][black long-short-dashed line]{Som}, 
\citet[][magenta short-dashed line; GALFORM]{Granato00}, and 
\citet[][red dot-long-dashed line]{Menci02} are shown.
The model of \citet{Som} is the `accelerated quiescent' model 
which was used in \cite{Som04}, and the model of \citet{Menci02}
is the `no-burst' model that was used in \citet{Poli} and 
\citet{Fontana03} for comparison with their data.
}
\label{sfhistory.eps}
\end{figure}

The scaling of the simulation results to the local estimate at $z=0$
leaves us with some concerns, because the relative value of $\Os$ does
indeed depend on this scaling. However, some kind of normalization 
is necessary to compare the results of different numerical models
on the same basis, because every numerical model has its input
parameters, and what we would like to focus in this paper is the 
{\it relative} speed of the development of $\Os$ from high redshift
to the present time.  The two TVD simulations in fact yield
different values of $\Os$ owing to cosmic variance and differences 
in the set of cosmological parameters adopted in the simulations: 
$\Os=0.0077$ and 0.0052 for N864L22 and N768L25, respectively.  
The normalization of $\Os$ at $z=0$ in Figure~\ref{sfhistory.eps} 
can be partially explained by the scaling of $\Os$ with the baryon 
mass density in the different simulations.  \citet{Her03} have shown 
that $\Os$ scales as $\Os \sim \Ob^{\,1.8}$ from theoretical 
arguments, and \citet{Gardner} found that the amount of cold 
gas and stars in their SPH simulations follow 
$\Os \propto \Ob^{1.0-2.0}$.

Using the cosmological parameters of the two TVD runs, the scaling
$\Os\propto \Ob^{\,1.8}$ gives the expected ratio 
$\Os({\rm N864L22})/ \Os({\rm N768L25})=1.7$. The actual corresponding 
ratio from the two simulations is $(0.0077/0.0052)=1.5$, which is a 
reasonable agreement, given other uncertainties, such as cosmic 
variance. It is expected that the H\&S model (blue long-dashed line in 
Figure~1) should yield the highest $\Os$ at $z=3$, because it is 
intended to remove the effects of limited resolution and cosmic 
variance. Since the results of N864L22 and N768L25 are not corrected 
for the limited box-size and resolution, it is natural that they lack 
the earliest star formation at $z>10$, resulting in a lower $\Os$ at 
$z=3$. This is also true of the SPH G6 run which by itself cannot 
resolve the entire starforming population at $z\gtsim 3$.  
Also, a comparison of the N864L22 and N768L25 results gives 
an idea of the level of cosmic variance for a volume of $\approx 
(25\himpc)^3$, and the deviation of the SPH G6 run from the H\&S model
is also a consequence of cosmic variance.

Figure~1b compares the results of H\&S theoretical model, 
semi-analytic models of \citet{Som}, \citet{Granato00}, \citet{Menci02},
and observations. The model shown for \citet{Som} is the updated 
`accelerated quiescent' model 
which was also used in \cite{Som04}, and follows the merger tree 
down to halos with circular velocity $30 ~\kms$. 
The result shown for \citet{Menci02} is the original model used by 
\citet{Poli} and \citet{Fontana03} for the comparison to their 
observational data. Recently \citet{Menci03} updated their model to 
include a starburst mode of star formation, which gives a result close
to that of GALFORM. The results of the semi-analytic models are somewhat 
higher than the observational estimates, but within the upper limit of 
\citet{Fontana03} data point at $z=3$. 
It is also clear that the predicted $\Os$ by the H\&S model is 
higher than that of the two semi-analytic models by about a factor 
of two. The discrepancy between the models and observations 
increases at $z\sim 1.5$, and is significant at more than 1-$\sigma$.

%%%%%%%%%%%%%%%%%%%%%%%%%%%%%%%%%%%%%%%%%%%%%%%%%%%%%%%%%%%%%%%%%%%%%%

\section{Cosmic Star Formation Rate at $0\ltsim z \ltsim 6$}
\label{sec:sfr}

\begin{figure}
\epsscale{1.1}
\plottwo{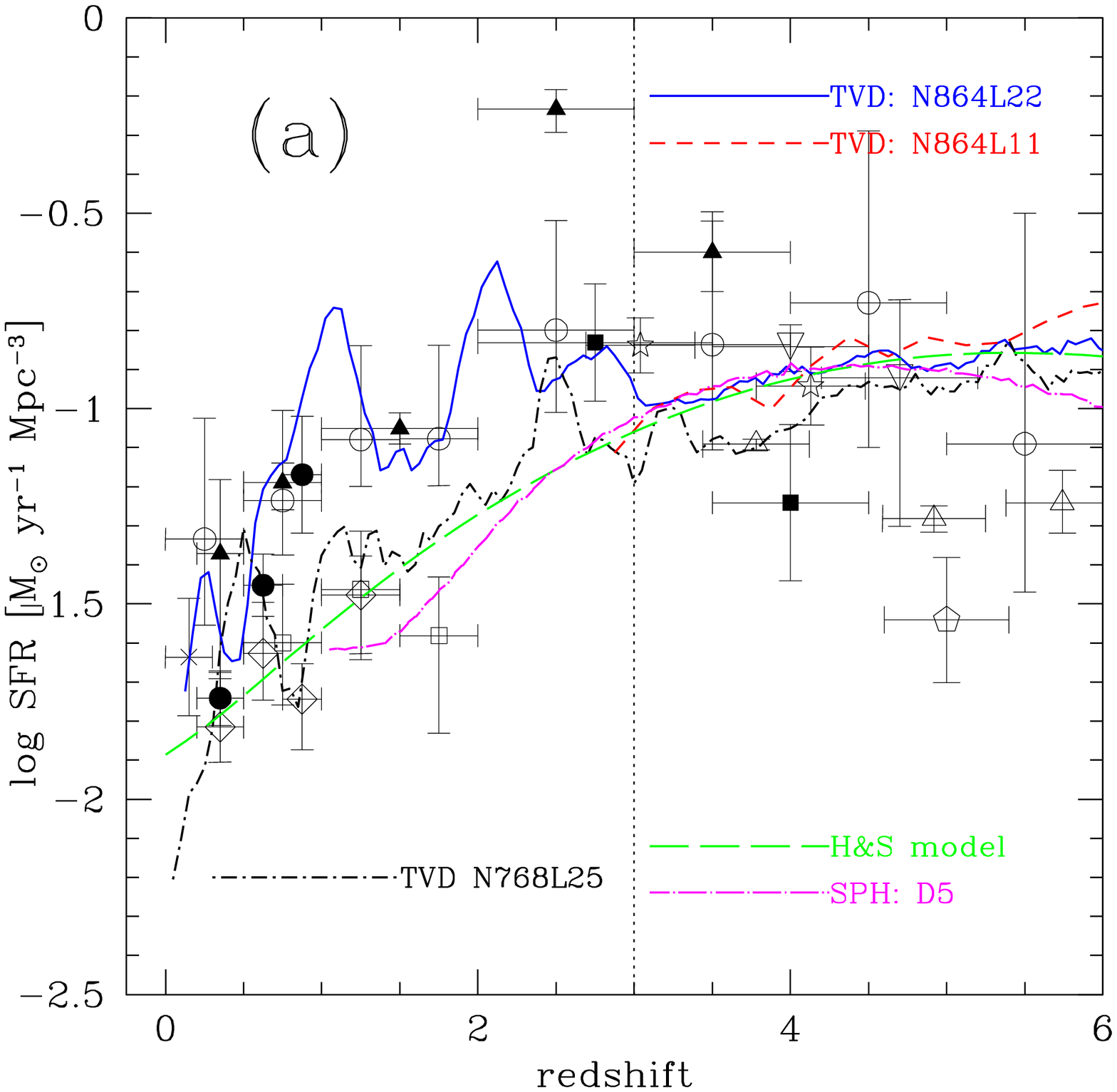}{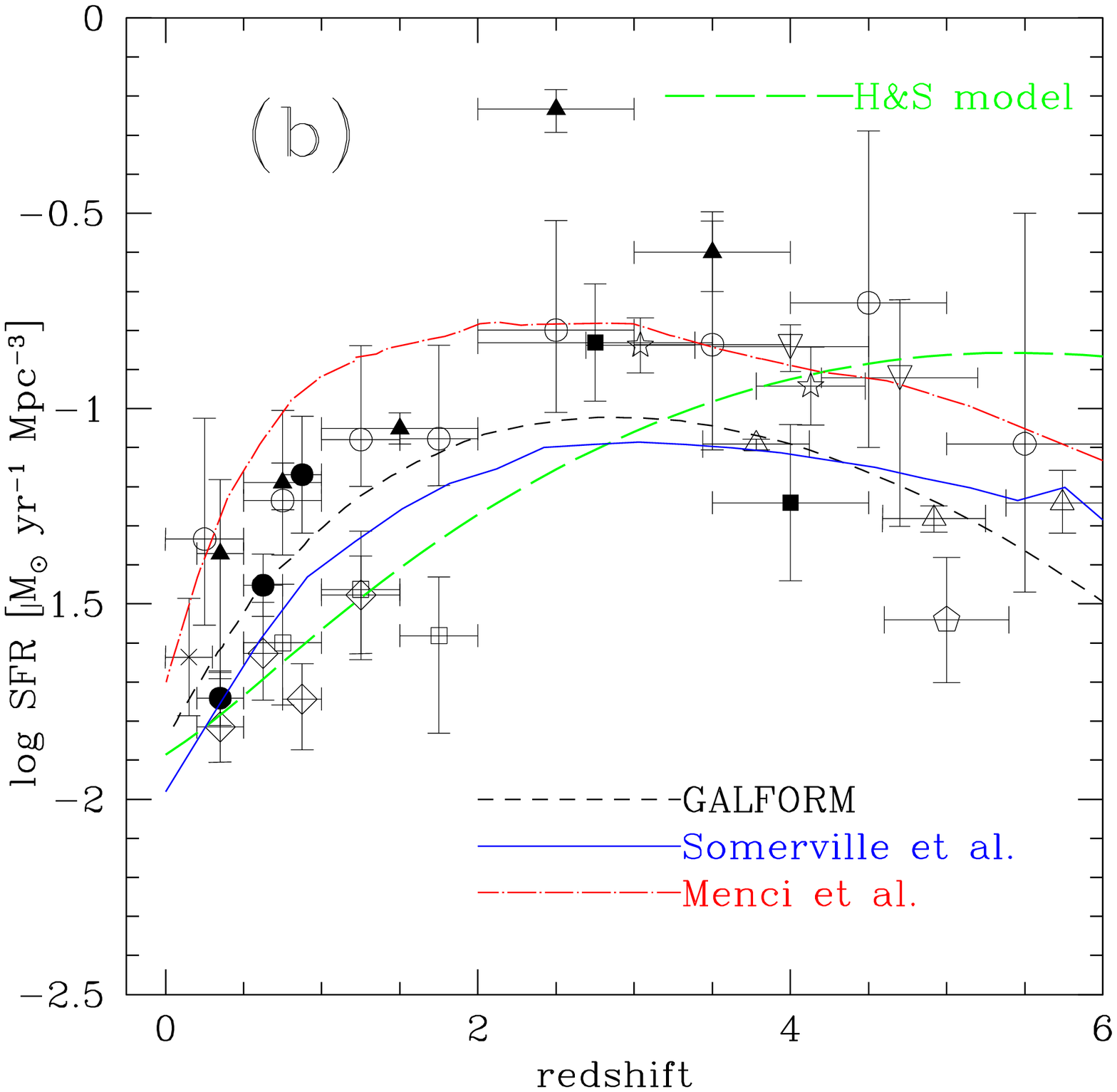}
\caption{Star formation rate density as a function of redshift.
{\it Panel (a)}: The lines correspond to TVD N864L22 (solid blue 
line), TVD N864L11 (red short-dashed line; stopped at $z=3$ owing to 
its small box-size), TVD N768L25 (black dot long-dashed line), 
SPH `D5'-run (magenta dot long-dashed line), SPH `G6'-run (cyan 
short-dash long-dash line), and the model of 
\citet[][green long-dashed line]{Her03} as given by 
equation~(\ref{eq:sfr}). The simulation results shown in this figure 
are extracted directly from the runs, without any further adjustments.
The result of the SPH Q5-run is not shown here because it follows the 
\citet{Her03} model line very closely at $z\geq 3$.
The sources of the (dust corrected) observational data points are: 
\citet[][filled circles]{Lilly96}, 
\citet[][filled squares]{Madau96, Madau97}, 
\citet[][open squares]{Connolly97}, 
\citet[][filled triangles]{Sawicki97}, \citet[][open cross]{Treyer98}, 
\citet[][open circles]{Pas98}, \citet[][open diamond]{Cowie99}, 
\citet[][open stars at $z=3,4$]{Steidel99}, 
\citet[][open pentagon at $z=5$]{Iwata}, 
\citet[][open inverted triangle at $z=4,5$]{Ouchi03b}, 
and \citet[][open triangles at $z=3-6$]{Giavalisco04}.
{\it Panel (b)}: H\&S model (green long-dashed line), semi-analytic 
models of \citet[][blue solid line]{Som}, \citet[][black short-dashed 
line]{Granato00}, and \citet[][]{Menci02} are compared with observations.
The model of \citet{Som} is the `accelerated quiescent' model 
which was used in \cite{Som04}, and the model of \citet{Menci02}
is the `no-burst' model that was used in \citet{Poli} and 
\citet{Fontana03} for comparison with their data. Both simulations 
and H\&S model has a peak of SFR at $z=5-7$, whereas the semi-analytic 
models have a peak at $z=2-4$.
}
\label{sfr.eps}
\end{figure}

In Figure~\ref{sfr.eps}, we show the cosmic SFR density as a function 
of redshift. Panel (a) shows the simulation results, and panel (b)
 shows the semi-analytic model results. The simulation results shown 
in this figure are extracted directly from the runs, without any 
further adjustments. The H\&S model that is shown in both panels has 
an approximate form as
\begin{equation}
\dot\rhos = \dot\rhos(0)
\frac{\chi^2}{1+\alpha(\chi-1)^3\exp{(\beta\chi^{7/4})}},
\label{eq:sfr}
\end{equation}
where $\chi(z) \equiv (H(z)/H_0)^{2/3}$. For a $\Lam$CDM universe 
with the star formation and feedback algorithm described by 
\citet{SH03a}, the parameters defining the SFR density take the 
values $\alpha=0.012$, $\beta=0.041$, and
$\dot\rhos(0)=0.013\,{\rm M_\odot ~yr^{-1}~Mpc^{-1}}$.
See Section~\ref{sec:conclusion} for further discussion on this 
formula.

The line types and data points are described in the caption.
The observationally estimated ultra-violet (UV) luminosity densities 
$\rho_{\rm UV}$ have been converted into the SFR by $\rho_{\rm UV} 
[{\rm erg~s^{-1}~Hz^{-1}~Mpc^{-3}}]= 8.0\times 10^{27} SFR [\Msun~\yr^{-1}~{\rm Mpc}^{-3}]$
\citep*{Madau98}. The SFR data points are corrected for dust 
extinction according to a prescription similar to that of 
\citet{Steidel99}: we assume the highly uncertain extinction correction 
factors to be 1.3 ($z<2$) and 4.0 ($z>2$), while \citet{Steidel99} used 
higher values 2.7 ($z<2$) and 4.7 ($z>2$).

It is important to note that each data point has been derived under 
different assumptions, because the faint-end slope of the luminosity 
function of galaxies at high redshift is not well constrained and 
adopting a steeper slope and integrating down to fainter magnitudes
would certainly yield a larger UV luminosity density. Here we describe
some of the high redshift data points. \citet[][open stars at 
$z=3, 4$]{Steidel99} derived their UV luminosity density by integrating 
the luminosity function with a faint-end slope of $\alpha=-1.6$ down 
to $0.1 L^*$. For their data points, we read off the SFR from Fig.9, 
corrected to our flat-$\Lam$ cosmology, and applied our extinction 
correction. \citet[][open triangles at $z=3-6$]{Giavalisco04} 
integrated the Schechter fit with a faint-end slope of $\alpha=-1.6$ 
down to $0.2 L^*_3$, where $L^*_3$ is the characteristic UV luminosity 
of LBGs at $z\sim 3$. \citet[][open pentagon at $z=5$]{Iwata} 
integrated their Schechter fit with $\alpha=-1.5$ in the magnitude 
range of $22.5< M_{\rm UV} - 5\log h< -20.0$. This integration range was 
chosen to match with the limiting magnitude of the $z\sim 4$ sample 
by \citet{Steidel99} in terms of absolute magnitude, therefore it is 
expected from the shape of the luminosity function that the resultant 
UV luminosity density at $z\sim 5$ is $0.56 - 0.69$ times of that 
at $z\sim 3$ in the same absolute magnitude range, depending on the 
choice of cosmology and the integration range. For the 
\citet[][inverted open triangles]{Ouchi03b} data points, we show
their total UV luminosity density case where the luminosity function
with a faint-end slope of $\alpha=-2.2$ is integrated down to 
$0.1 L^*$. 

An important theoretical result here is that all the simulation results 
and the H\&S model peak at $z\geq 5$, and not at lower redshifts, 
as is often found in semi-analytic models of galaxy formation 
\citep[e.g.][see panel (b)]{Baugh, Kau99, Cole00, Som}. 
Observationally, it is an unsettled problem whether the SFR levels 
off at high-redshift \citep[e.g.][]{Steidel99} or still increases beyond 
$z>3$ \citep{Lanzetta}. In particular, \citet{Lanzetta} stress the 
importance of the cosmological surface dimming effect and argue that 
previous measurements have missed a significant fraction of the 
ultraviolet luminosity density of the universe at $z\geq 2$.
The scatter in the data points at $z>3$ is still large, and it is
not possible at this time to determine the trend in SFR at these 
redshifts. However, our point is a theoretical one, and does not 
depend on the distribution of data points as we detail in 
Section~\ref{sec:conclusion}.

In passing, we note that the absolute values of $\Os$ at $z=0$
computed by integrating the SFR curves shown in Fig.2b are 
0.0037, 0.0031, 0.0041, and 0.0090 for the models by \citet{Her03}, 
\citet{Som}, \citet{Granato00}, and \cite{Menci02}, respectively.

The result of TVD N864L22 at $z<1$ is slightly higher than most of
the observational data points and has two large bumps at $z\sim 1.0$
and 2.0.  These two peaks presumably owe to major merger events
taking place in the simulation, and are a consequence of cosmic
variance in the relatively small box utilized. If we had utilized 
a larger box, then we would expect the curve to be smooth. Clearly 
the box-size of $\Lbox=20-30\himpc$ is not large enough to accurately 
model the volume averaged quantities in the universe at $z<1$.

%%%%%%%%%%%%%%%%%%%%%%%%%%%%%%%%%%%%%%%%%%%%%%%%%%%%%%%%%%%%%%%%%%%%%%

\section{Discussion \& Conclusions}
\label{sec:conclusion}

We have shown that two independent different types of numerical 
hydrodynamic simulations both predict that the cosmic star
formation rate density should peak at $z\geq 5$, and that
this relatively early peak in the SFR leads to a more rapid 
development of the stellar mass density than current observational 
estimates.  When all the results are scaled to the local value 
at $z=0$, the stellar mass density at $z=3$ in the simulations is 
larger than observed values by more than a factor of two.  
We also showed in Figure~1b that the semi-analytic models of 
\citet{Som}, \citet{Granato00}, and \citet{Menci02} predict larger 
$\Os$ than current observational estimates, but within the upper 
error bar of the \citet{Fontana03} data point at $z=3$. 
In particular, the H\&S model predicts higher $\Os$ than 
those semi-analytic models by about a factor of two at $z=3$.
This comes from the fact that both the simulations and H\&S model 
has a peak of SFR at $z=5-7$, whereas the semi-analytic models 
have a peak at $z=2-4$. 

The high $\Os$ predicted by our simulations and the model of
\citet{Her03} suggest that current observations are missing 
nearly half of the total stellar mass density in the universe at 
high redshifts. This missing stellar mass could be hidden in a 
population of red galaxies that have not been detected previously 
in the optical ground-based data. In fact, \citet{Franx} and 
\citet{Daddi04} find such a population of red galaxies at 
$z\geq 2$ in HDF-South that has a volume density half that of LBGs
at $z=3$ and a stellar mass density comparable to that of LBGs. 
However, the data points by \citet{Rudnick03} include this 
red galaxy population. Therefore the cosmic variance might be a 
stronger cause for the underestimate of $\Os$ by the current 
observations. This is hinted by the absence of this red population 
in HDF-North. If these red 
populations are strongly clustered as suggested by \citet{Daddi03}, 
observations with small fields-of-view could easily miss them. 
The stellar mass density in the simulations and observations agree 
reasonably well at $z<1$ where the rate of increase is much slower 
than at higher redshift. In addition, the dust extinction correction
may seriously underestimate the fraction of early star formation
that is heavily obscured. If we only had optical UV observations
of nearby starburst galaxies -- rather than the full spectrum 
extending to the submillimeter -- we would greatly underestimate
star formation in these systems \citep[e.g.][]{Hughes98, Meurer99, 
Barger00, Calzetti00, Takagi03}. 
Future studies of starforming galaxies
in the far-infrared wavelengths \citep[e.g.][]{Kennicutt03} using 
the {\it Spitzer Space Telescope}, in the submillimeter 
\citep[e.g.][]{Tecza04}, in the millimeter 
\citep[e.g.][]{Bekki00} using the {\it Atacama Large Millimeter 
Array} (ALMA), and in X-rays \citep[e.g.][]{Norman04} using the 
{\it Chandra X-ray Observatory} would enable us to constrain the 
nature of starburst galaxies better, and estimate the intrinsic
SFR more accurately.

It is reassuring that the two different sets of simulations, which use
very different hydrodynamic methods (i.e. Eulerian TVD and SPH), both
give a star formation history that peaks at $z\geq 5$.  As argued by
\citet{Her03} this is to be expected, because the evolution of the
cosmic SFR is driven mainly by a competition between gravity and the
expansion of the Universe, with a weaker dependence on the details of
star formation and feedback. The form of equation~(\ref{eq:sfr}) given 
in Section~\ref{sec:sfr} can be understood as follows. At high 
redshifts, when the cooling time is short, star formation is limited 
primarily by the gravitational growth of halos, which is independent 
of the dissipative gas dynamics. Thus, the parameter $\beta$ and the 
factor $\chi^{7/4}$ in the exponential of equation~(\ref{eq:sfr}) are 
determined by the form of the matter power spectrum, and the 
description of star formation and feedback enter into $\dot\rhos(z)$ 
only logarithmically. At low redshifts, the supply of star forming 
gas is limited by the expansion rate of the Universe, fixing the 
dependence of $\dot\rhos(z)$ on $\chi$ as $z\rightarrow 0$. The 
explicit influence of the prescription for star formation and 
feedback is again subdominant and mainly affects the values of the 
normalization parameters $\alpha$ and $\dot\rhos(0)$.

For these reasons, equation~(\ref{eq:sfr}) can be generalized 
straightforwardly to other cosmologies and to include other physics
\citep[e.g.][]{Yoshida03}.  Moreover, as shown in e.g. figure 6 of 
\citet{Her03}, the fact that the SFR density is regulated mainly by 
the competition between gravity and the expansion of the Universe 
means that the peak in $\dot\rhos(z)$ should lie at $z\approx 5$, 
unless an implausible value is adopted for the parameters governing 
star formation. Hence, it is not surprising that the various sets 
of simulations should be consistent, with residual differences owing
to details in the cosmology and, most important, cosmic variance and
resolution limitations.  We plan to investigate these issues in
the future using the algorithms described here, as well as adaptive 
mesh refinement codes.

For now, the agreement between our different numerical approaches
supports the general arguments made by Hernquist \& Springel (2003),
that the SFR density should peak at $z\geq 5$, mostly independent of 
the details of star formation and feedback.  As we have demonstrated 
here, such an early peak in the cosmic star formation history 
yields $\Os$ that clearly exceeds current observational estimates 
and the results of semi-analytic models, suggesting that most of 
the stars in the universe at $z\gtsim 3$ are ``missing.''

%%%%%%%%%%%%%%%%%%%%%%%%%%%%%%%%%%%%%%%%%%%%%%%%%%%%%%%%%%%%%%%%%%%%%%

\acknowledgments 

We thank Rachel Somerville, Carlton Baugh, Shaun Cole, and 
Nicola Menci for providing the result of their semi-analytic model. 
This work was supported in part by NSF grants ACI 96-19019, 
AST 98-02568, AST 99-00877, AST 00-71019, and AST-0206299, and 
NASA ATP grant NAG5-12140 and NAG5-13292, NAG5-13381.
The SPH simulations were performed at the Center for Parallel
Astrophysical Computing at Harvard-Smithsonian Center for
Astrophysics. The TVD simulations were performed at the National
Center for Supercomputing Applications (NCSA).

%%%%%%%%%%%%%%%%%%%%%%%%%%%%%%%%%%%%%%%%%%%%%%%%%%%%%%%%%%%%%%%%%%%%%%

%%%%%%%%%%%%%%%%%%%%%%%%%%%%%%%%%%%%%%%%%%%%%%%%%%%%%%%%%%%%%%%%%%%%%%

\begin{deluxetable}{cccccc}  
%\tablenum{\tabnum}
\tablecolumns{6}  
\tablewidth{0pc}  
\tablecaption{Simulations}
\tablehead{
\colhead{Run} & \colhead{$\Lbox$ [$\himpc$]} & \colhead{$N_{\rm mesh/ptcl}$} & \colhead{$m_{\rm DM}$ [$\himsun$]}  & \colhead{$m_{\rm gas}$ [$\himsun$]} & \colhead{$\Del\ell$ [$\hikpc$]}   
}
\startdata
TVD: N864L11$^a$ & 11.0 & $864^3$ & $1.1\times 10^6$ & $2.7\times 10^4$ & 12.7\cr
TVD: N864L22$^a$ & 22.0 & $864^3$ & $8.9\times 10^6$ & $2.2\times 10^5$ & 25.5\cr
TVD: N768L25$^b$ & 25.0 & $768^3$ & $2.0\times 10^7$ & $3.4\times 10^5$ & 32.6\cr
SPH: Q5$^c$ & 10.0  & $324^3$ & $2.1\times 10^6$ & $3.3\times 10^5$ & 1.2\cr
SPH: D5$^c$ & 33.75 & $324^3$ & $8.2\times 10^7$ & $1.3\times 10^7$ & 4.2\cr
SPH: G6$^c$ & 100.0 & $486^3$ & $6.3\times 10^8$ & $9.7\times 10^7$ & 5.3\cr
\enddata
\tablecomments{
Parameters of some of the simulations on which this study is based.
The quantities listed are as follows:
$\Lbox$ is the simulation box-size, $N_{\rm mesh/ptcl}$ is the number of
the hydrodynamic mesh points for TVD or the number of gas particles for 
SPH, $m_{\rm DM}$ is the dark matter particle mass, $m_{\rm gas}$ is the 
mass of the baryonic fluid elements in a grid cell for TVD or the 
masses of the gas particles in the SPH simulations. Note that the TVD uses $432^3$ ($384^3$) dark matter particles for N864 (N768) runs. $\Del\ell$ is the 
size of the resolution element (cell size in TVD and gravitational 
softening length in SPH in comoving coordinates; for proper distances,
divide by $1+z$).
The upper indices on the run names correspond to the following sets
of cosmological parameters: $(\Om, \Ol, \Ob, h, n, \sigma_8)=
(0.29, 0.71, 0.047, 0.7, 1.0, 0.85)$ for (a), $(0.3, 0.7, 0.035, 0.67, 
1.0, 0.9)$ for (b), and $(0.3, 0.7, 0.04, 0.7, 1.0, 0.9)$ for (c).
}
\label{table:simulation}
\end{deluxetable}

\end{document}